# Spectroscopic and Photometric Study of the new δ Scuti Star ASAS J063309+1810.8

Mohamed I. Nouh, Mohamed Abdel-Sabour, Ahmed Shokry, Gamal M. Hamed, Diaa A. Fouda and Ali Takey

Astronomy Department, National Research Institute of Astronomy and Geophysics, 11421 Helwan, Cairo, Egypt

mohamed.nouh@nriag.sci.eg

**Abstract:** We present BVR observations and low-resolution spectra collected by the Kottamia Astronomical Observatory 1.88 m telescope (KAO) for the new pulsating star ASAS J063309+1810.8 (hereafter it will be called ASAS06+18). The photometric analysis revealed that the star is a δ Scuti star with low amplitude (a=0.054-0.099 in V mag.) and a short period (102.604 min). Fourier analysis of the light curves reveals the fundamental mode with two harmonics. The photometric analysis yielded a new value of the updated frequency of 13.0035232 cd$^{-1}$ with an amplitude of 49.93 mmag at phases 0.326 and S/N 21.75 and to two frequencies (20.2099237cd$^{-1}$, 5.9130945cd$^{-1}$). Given the available data, 37 new times of maximum light are presented, and an updated ephemeris for the star and its O-C data. Assuming its period decreases and changes smoothly, a new value of $(1/P)\,dP/dt$ is determined. We calculated the effective temperature and surface gravity as $T_{eff} = 7125 \pm 250$ K and $\log g = 4.0 \pm 0.2$ dex from model atmosphere analysis of the star's spectra at different phases. The bolometric magnitude $M_{bol}$=2.798±0.016, radius R=1.577±0.077$R_\odot$, luminosity L=5.714±1.066$L_\odot$, the mass is M=1.595 $M_\odot$, and pulsation constant Q=0$^m$.0338±0.0003. The star's locations in the evolutionary mass-luminosity and mass-radius relationships are discussed.

## 1. Introduction

The δ Scuti stars have short periods and small amplitude variations. They also have radial and nonradial p-mode pulsations. Their luminosities range from V to III (sub-giant), and spectral types range from (nearly) F8 to A2, Netzel et al. (2022). Furthermore, these stars are in a series of evolutionary states and can be found closer to the main sequence within the instability strip.

These pulsators have a variety of subtypes and behaviours that are worth describing in order to locate some trails through the δ Scuti jungle. The δ Scuti stars are classified into sub-types: stars with high amplitude ΔV > 0$^m$.3 (HADS) and stars with a low amplitude variable with

ΔV < 0$^m$.1 (LADS), Rodriguez and Breger (2001). Another subgroup of δ Scuti stars is the SX Phe variables of the Pop. II and old disk population. They are typically found in the galaxy's outer regions, known as the galactic halo. Their luminosity changes over 1-2 hours. They're fascinating because, despite their age, they haven't converted into white dwarfs, as one would assume. SX Phe spectrum classifications are in the A2-F5 range, also displaying light amplitudes from 0.003 to 0.9 mag in V color. Most SX Phe stars are also HADS, but not vice versa (Berger, 2000).

For decades, astronomers have known that the LADS has a broad spectrum of nonradial modes, complicated light variability, multi-periodicity, phase, and amplitude fluctuations. Previously, it was assumed that HADS were classical radial pulsating stars, generally in a mono periodic double mode, but always pulsating in radial modes (Kjurkchieva et al., 2013). However, many HADS have recently been discovered to be multi-periodic variables with nonradial as well as radial pulsations, Zhou (2002), and Poretti (2003).

The new pulsating star ASAS06+18 (ASAS J063309+1810.8, UCAC4 541-029048, TIC 54609784) is classified as a δ Scuti star by Pojmanski et al. (2005) with spectral type F0, period of pulsation p=0.071253 d and metallicity [Fe/H] = -0.20±0.09. Qian et al. (2018) analyzed LAMOST spectroscopic observations over one night and calculated the effective temperature and the surface gravity as $T_{eff} = 7279 \pm 100\,K$ and $\log g = 3.89 \pm 0.13$ dex, respectively.

The present paper examines the photometric and spectroscopic observations of the new δ Scuti star ASAS06+18 obtained at the Kottamia astronomical observatory (KAO, Egypt). Light curve analysis of the BVR colors and model atmosphere analysis for the spectroscopic observations will be performed to determine the absolute physical parameters of the star. The paper is organized as follows: Section 2 represents the photometric analysis, Section 3 discusses the spectral analysis, Section 4 is devoted to the determination of the physical parameters and evolution state of the star, and Section 5 summarizes the results.

## 2. Photometric Analysis

### 2.1. Observations and Data Reduction

On four nights, February 12$^{th}$, 13$^{th}$, 14$^{th}$, and 18$^{th}$, 2021, the star ASAS06+18 was observed in the B, V, and R bands at KAO using the Kottamia Faint Imaging Spectropolarimeter (KFISP) installed on the Cassegrain focus of Kottamia Astronomical Observatory (KAO, Azzam et al. 2021). The raw CCD images were subjected to bias subtraction and flat-field correction without dark subtraction, which was already negligible (the CCD camera was stably working at a very low temperature of -120 C). P. Nosal[1] observed the star in the V filter using NWT 150/600, Canon 20 D, and MPCC mk.III instruments at Perek observatory (Czech Republic).

Figure 1 depicts the field of observations for the star ASAS06+18, where V denotes the variable star, C1 denotes the comparison star, and C2 denotes the check star. In Table 1, we listed the basic information of the three stars V, C1, and C2. Also, we listed some of the available colors from different databases. Table 2 contains the observed data points for each filter, the exposure time, and the total time for each night.

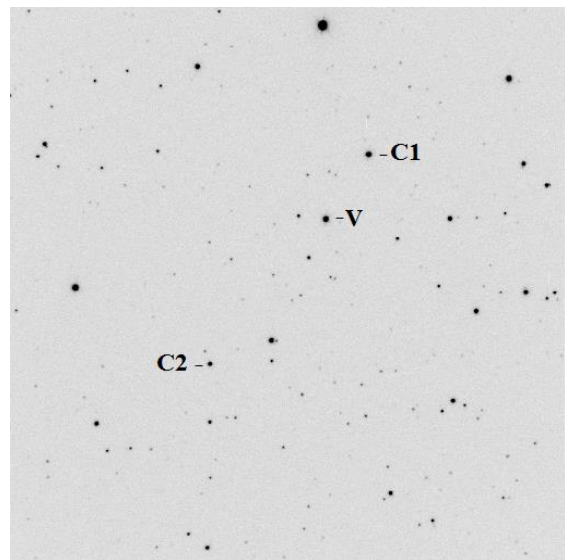

Figure 1. The Chart of the variable (V), comparison (C1), and check (C2) stars.

Table 1. Coordinates, magnitudes, and a color index of the Variable and Comparison stars.

| Star | Variable (V) | Comparison (C1) | Check (C2) |
|---|---|---|---|
| ID | ASAS J063309+1810.8 ASASSN-V J063309.02+181045.2 UCAC4 541-029048 | 2MASS 06331058+1807209 | 2MASS 06331599+1812572 |
| RA(J2000) "h:m:s" | $06^h\ 33^m\ 09.02^s$ | $06^h\ 33^m\ 10.58^s$ | $06^h\ 33^m\ 15.99^s$ |
| DEC(J2000) "d:m:s" | +18° 10′ 45.5″ | +18° 07′ 20.9″ | +18° 12′ 57.27″ |
| J | 11.112±0.018* | 15.438±0.058 | 12.349±0.021 |
| J-H | 0.172±0.0240* | 0.508±0.0450 | 0.316±0.0130 |
| J-K | 0.21* | | |
| B-V | 0.33** | | |

(https://www.aavso.org/vsx/index.php?view=detail.top&oid=80253).
*2MASS (2MASS All-Sky Catalog of Point Sources).
**APASS-DR9 (AAVSO Photometric All Sky Survey).

[1] http://var2.astro.cz/obslog.php?obs_id=1&star=ASAS%20J063309%2B1810.8

Table 2: Log table of the photometric observation of the star ASAS06+18.

| Date/filter | No. of data points | | | | Exp. time (sec.) |
|---|---|---|---|---|---|
| | 12/02/2021 | 13/02/2021 | 14/02/2021 | 18/03/2021 | |
| B | 35 | 38 | 0 | 0 | 100 |
| V | 38 | 39 | 32 | 32 | 30 |
| R | 29 | 36 | 26 | 35 | 10 |
| Total time (min) | 130 | 155 | 70 | 180 | |

## 2.2. Light Curve and Pulsation Analysis

The MuniWin v.1.1.26 software (Hroch, 1998) was used for aperture photometry, where we obtained the differential magnitude of the variable star. The light curves were generated using the magnitude differences between the system and the comparison star. Photometric accuracy was estimated using the standard deviations of the magnitude differences between the check star and the comparison star, with a mean value of 0.003 m in ideal observing conditions and 0.013 m in poor conditions. We adjusted it using the data fit light curves for each month, despite slight zero-point adjustments (assuming the frequencies are stable in one month). Figure 2 exhibits KAO observations from the Johnson BVR as well as V-mag measurements from the Perek observatory.

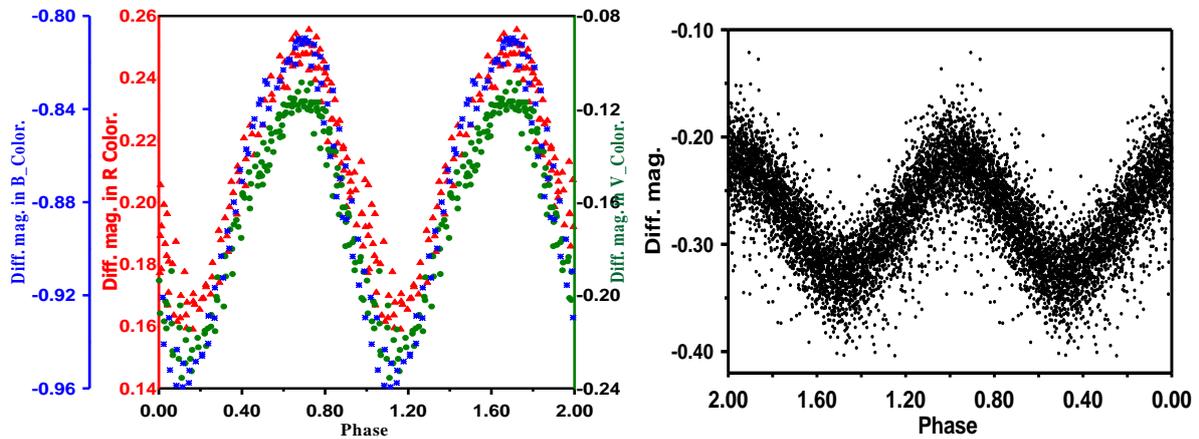

Figure 2. The right panel shows BVR phased light curves (for KAO Observations) for ASAS06+18, while the left panel shows the phased light curve from Perek Observations, both with the original period *P*=0.071253 d.

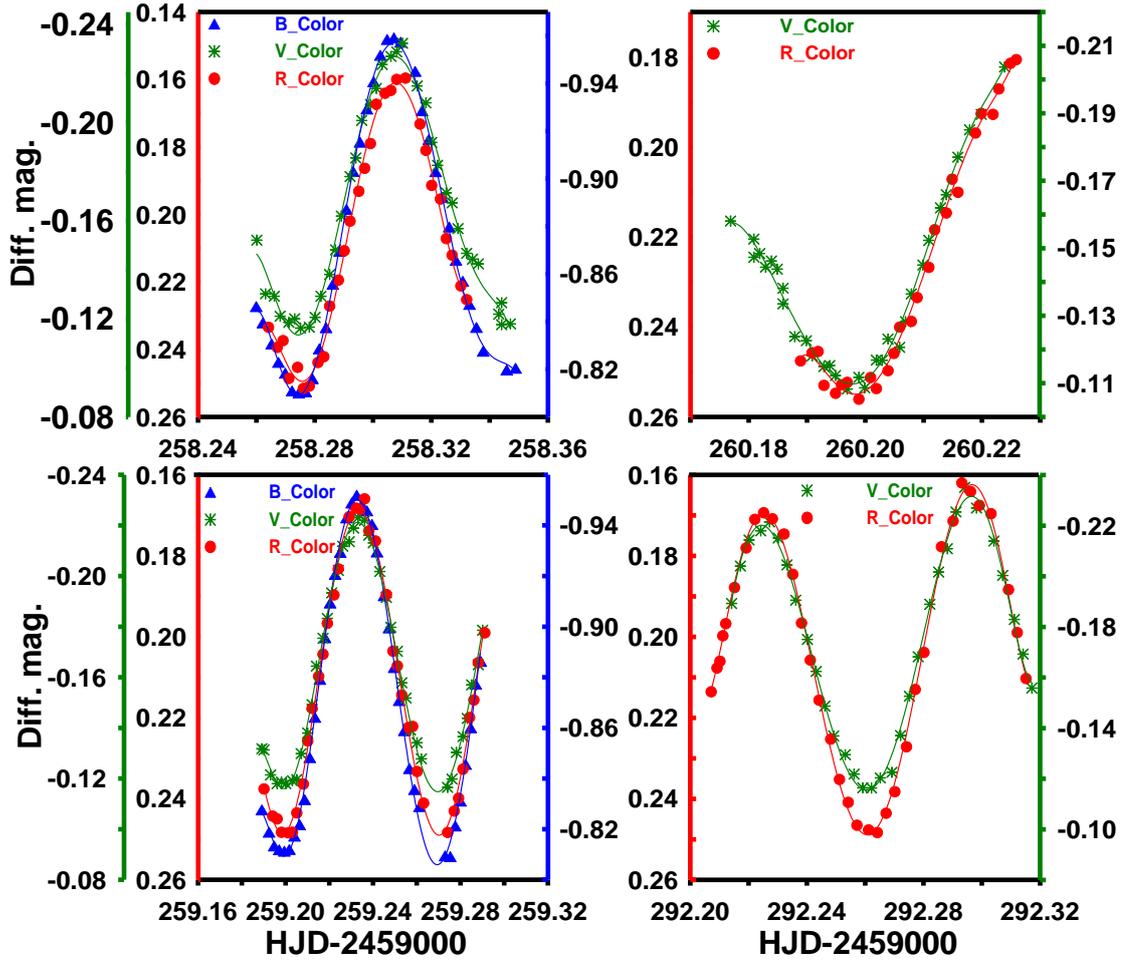

Figure3. Differential magnitude in BVR_Color light curves with the fitted line of the pulsating variable ASAS06+18.

### 2.2.1. Frequency Analysis

Peranso[2] and PERIOD04 (Lenz and Breger, 2005) codes were used to analyse the frequency of the light curves of the pulsating star ASAS06+18. They both looked for prominent peaks in the amplitude spectra using Fourier transformations of light curves. After computing the initial frequency, we calculate the "periodogram" by fitting a sinusoid to the Period04 period and then subtracting the sinusoid from the original magnitude (pre-whitening). The periodogram is then calculated again, but the initial frequency is absent; therefore, the periodogram's greatest peak reflects the following frequency. This technique is repeated as many times as necessary to hunt for more peaks until no more peaks can be located.

The results of the fitting Fourier transformations equations on the new data (points in Fig.3) are presented in Table 2, i.e., frequency, amplitude-phase, and signal-to-noise ratio (S/N). As listed in the last column of Table 2, the frequencies with a signal-to-noise ratio (S/N) greater than 4.0 were considered (Breger et al. 1993). The power spectrum of ASAS06+18 is depicted in Figure 4. Given the amplitude values in the second column of Table 2 and the power spectrum in Figure 4, it can be concluded that the fundamental frequency of ASAS06+18 is 13.0035232 cd$^{-1}$, and two additional frequencies were resolved in the residual spectrum after prewhitening the fundamental frequency, as we can see in Table 3.

[2]www.cbabelgium.com/peranso/

The δ Scuti stars are expected to have a frequency range of 3-80 c/d, with frequencies less than 3 c/d caused by atmospheric effects or observational errors. The error for each value was calculated using the sum of the squared residuals generated from a multi-parameter least-squares fit of sinusoidal functions. The frequency spectra, Fourier fits on the observational sites for all sets of data, and the spectral window of each star are depicted in Figure 4.

.

Table 3: The frequency analysis results.

|  | Frequency (cd$^{-1}$) | Amplitude (mmag) | Phase (0-1) | S/N |
|---|---|---|---|---|
| $f_1$ | 13.003523±0.000160 | 49.93±0.50 | 0.326±0.002 | 21.75 |
| $f_2$ | 20.209923±0.001047 | 8.39±0.30 | 0.317±0.010 | 5.372 |
| $f_3$ | 5.9130945±0.001961 | 4.17±0.50 | 0.062±0.019 | 4.705 |

Zeropoint:  -0.1676529
Residuals:   0.0051355

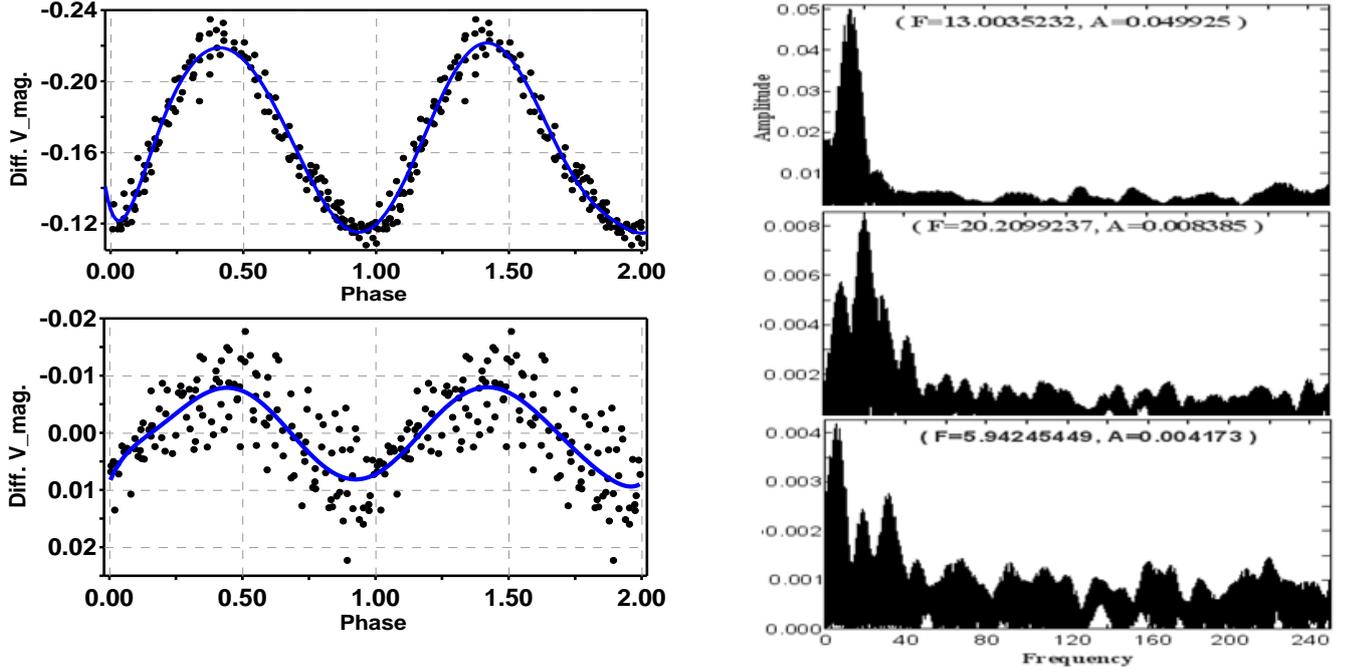

Figure 4. The Left panel (upper and lower) are the V-color data phased in two frequencies ($f_1$, upper; $f_2$, bottom), while the right panel is the amplitude spectra showing the pre-whitening process and resolved frequencies.
.

### 2.2.2. O-C Curve and Period Change

We utilized Hertzsprung (1928) approach to establish the time of lowest or maximum brightness, which was then used to build the O-C to look for a shift in the period of the star. We employed the approach proposed by Abdel-Sabour et al. (2020) to derive O-C differences from a calculated linear ephemeris. We create a maximum light reference time using existing photoelectric data, and the selected pulsation period is based on the KAO Observatory's recent observations of the star. KAO photometry in V color was utilized as a benchmark for matching data from the reference set's other epochs in phase and magnitude. Furthermore, the revised ephemeris was validated using all available ASAS photometry measurements for the star to guarantee its consistency throughout all observational epochs. The resultant times of maximum light ephemeris are given by: $HJD_{MAX} = 2452622.0160 + 0.071253065 E$.

Table 4 lists the 37 new times of maximum light we've determined. Unfortunately, we have no additional observations of this star, only a few light curves from ASAS and KAO. As shown in Figure 4, we used all available observations to construct the O-C. The following relationship gives the least-square fit by the parabolic elements of the star under consideration:

$$HJD_{max} = M_0 + PE + QE^2,$$

Where $M_0$ denotes a new epoch, P denotes a new period, and Q denotes the period change values ($dP/dt$) in seconds per year and calculated as $dP/dt = (2Q/P)\ 365.25 \times 24 \times 60 \times 60$. We fit the O–C residuals with a second-order polynomial least-squares fit with a root mean square of $R^2 = 0.954$ and obtained;

$$HJD_{MAX} = -2.011370(99) \times 10^4 + 1.63870(84) \times 10^{-2}\ E - 3.33793(42) \times 10^{-9}\ E^2,$$

In O-C data, the parabolic trend represents a regular period of decrease or increase. After constructing the O-C diagram, we found the O-C trend is decreasing period at a rate of $dP/dt = 2.9567 \pm 0.005\ s/yr$ (or $\frac{1}{P}(dP/dt) = 41.49579471\ yr^{-1}$), the standard deviation of the residuals of parabolic fit to the O - C values $0^d.005$ with a correlation coefficient of 0.95.

Table 3. New 37 times of maximum light, new Epoch, O-C, Number of observations, and the data source.

| HJD | Epoch | O-C | No. of Obs. | Source |
|---|---|---|---|---|
| 2451537.6043 | -15219.0 | 0.0103 | 81 | NSVS |
| 2452685.4994 | 891.0 | 0.0638 | 45 | ASAS |
| 2453014.8426 | 5513.0 | 0.0641 | 56 | ASAS |
| 2453350.7513 | 10227.0 | 0.0656 | 93 | ASAS |
| 2453728.6376 | 15531.0 | 0.0691 | 30 | ASAS |
| 2454398.5155 | 24932.0 | 0.0613 | 61 | ASAS |
| 2454855.7999 | 31350.0 | 0.0659 | 40 | ASAS |
| 2456480.3685 | 54150.0 | 0.0524 | 21 | APASS |
| 2457575.6958 | 69522.0 | 0.0342 | 111 | APASS |
| 2457692.6392 | 71164.0 | 0.0321 | 71 | Perek Obs. |
| 2457706.5631 | 71359.0 | 0.0339 | 134 | Perek Obs. |
| 2457715.5529 | 71485.0 | 0.0328 | 176 | Perek Obs. |
| 2457721.5728 | 71570.0 | 0.0360 | 156 | Perek Obs. |
| 2457725.5089 | 71625.0 | 0.0346 | 196 | Perek Obs. |
| 2457726.5047 | 71639.0 | 0.0342 | 238 | Perek Obs. |
| 2457727.4994 | 71653.0 | 0.0349 | 233 | Perek Obs. |
| 2457728.5223 | 71667.0 | 0.0317 | 338 | Perek Obs. |
| 2457741.5192 | 71850.0 | 0.0346 | 216 | Perek Obs. |
| 2457742.3700 | 71862.0 | 0.0349 | 34 | Perek Obs. |
| 2457751.4651 | 71989.0 | 0.0392 | 171 | Perek Obs. |
| 2457752.4332 | 72003.0 | 0.0349 | 233 | Perek Obs. |
| 2457753.4638 | 72017.0 | 0.0339 | 164 | Perek Obs. |
| 2457754.4213 | 72031.0 | 0.0331 | 101 | Perek Obs. |
| 2457759.4136 | 72101.0 | 0.0356 | 235 | Perek Obs. |
| 2457760.4123 | 72115.0 | 0.0342 | 233 | Perek Obs. |
| 2457762.4111 | 72143.0 | 0.0356 | 218 | Perek Obs. |
| 2457771.3902 | 72269.0 | 0.0335 | 212 | Perek Obs. |
| 2457772.3895 | 72283.0 | 0.0331 | 208 | Perek Obs. |
| 2457773.3812 | 72297.0 | 0.0335 | 218 | Perek Obs. |
| 2457774.3968 | 72311.0 | 0.0346 | 247 | Perek Obs. |
| 2457775.3761 | 72325.0 | 0.0335 | 219 | Perek Obs. |
| 2457777.3436 | 72352.0 | 0.0349 | 166 | Perek Obs. |
| 2457778.3028 | 72366.0 | 0.0339 | 98 | Perek Obs. |
| 2459258.2994 | 93137.0 | 0.0014 | 36 | KAO |
| 2459259.2325 | 93150.0 | 0.0000 | 40 | KAO |
| 2459260.1944 | 93164.0 | 0.0007 | 32 | KAO |
| 2459292.2633 | 93614.0 | 0.0043 | 32 | KAO |

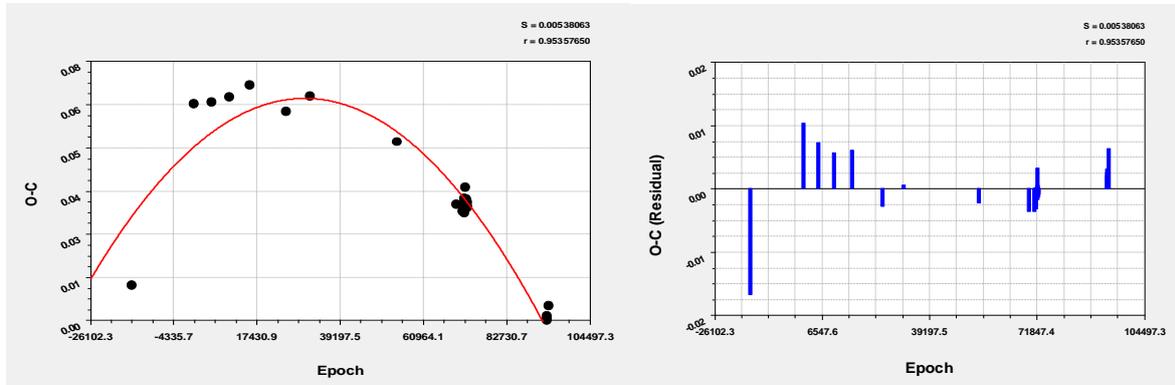

Figure5: O-C data points fitted with a quadratic (left panel) and the residuals of the quadratic fitting (right panel).

## 3. Model Atmosphere Analysis

KFISP observed ASAS06+18 with two grisms encompassing the spectral regions $\lambda\lambda 3360\text{-}5870\,\overset{o}{A}$ and $\lambda\lambda 5300\text{-}9180\,\overset{o}{A}$, with spectral resolutions of $1025\,\overset{o}{A}$ in blue and $1133\,\overset{o}{A}$ in red. The spectra were taken over two nights (7,8-Jan-2022). To reduce the data, the Astropy-affiliated software CCDPROC was utilized (Craig et al., 2017). We utilized van Dokkum and Pieter (2001) LACosmic technique to remove cosmic rays from the image using Astro-SCRAPPY (McCully et al., 2018). To extract the spectra and calibrate the wavelength, a custom Python method was employed with slight modifications (Khamitov et al., 2020). Flux calibration for the spectra was performed using IRAF procedures. The derived function of the specutils snr (a module for analysis tools dealing with uncertainties or error analysis in spectra) was used to compute the signal-to-noise ratio (Earl et al., 2021). Figure 6 depicts the ASAS06+18 spectra. The blue region is represented by the upper panel, while the lower panel represents the red zone.

The SPECTRUM code synthesizes LTE spectra (for the range $\lambda\lambda$ 1500–8000 $\overset{o}{A}$ ), Gray (1992). SPECTRUM calculates the columns for mass depth points, temperatures, and total pressure at each process stage using a system of seven nonlinear equilibrium equations. For LTE calculations, we utilized ATLAS9 grids (Kurucz, 1995) with solar metallicity, a microturbulent velocity of 2 km/s, and a mixing length to scale height ratio of 1.25. We constructed a small grid of synthetic spectra from LTE model atmospheres over the effective temperature range of 250 K for the spectroscopic study. Figure 7 shows the normalized synthetic spectra for three different models.

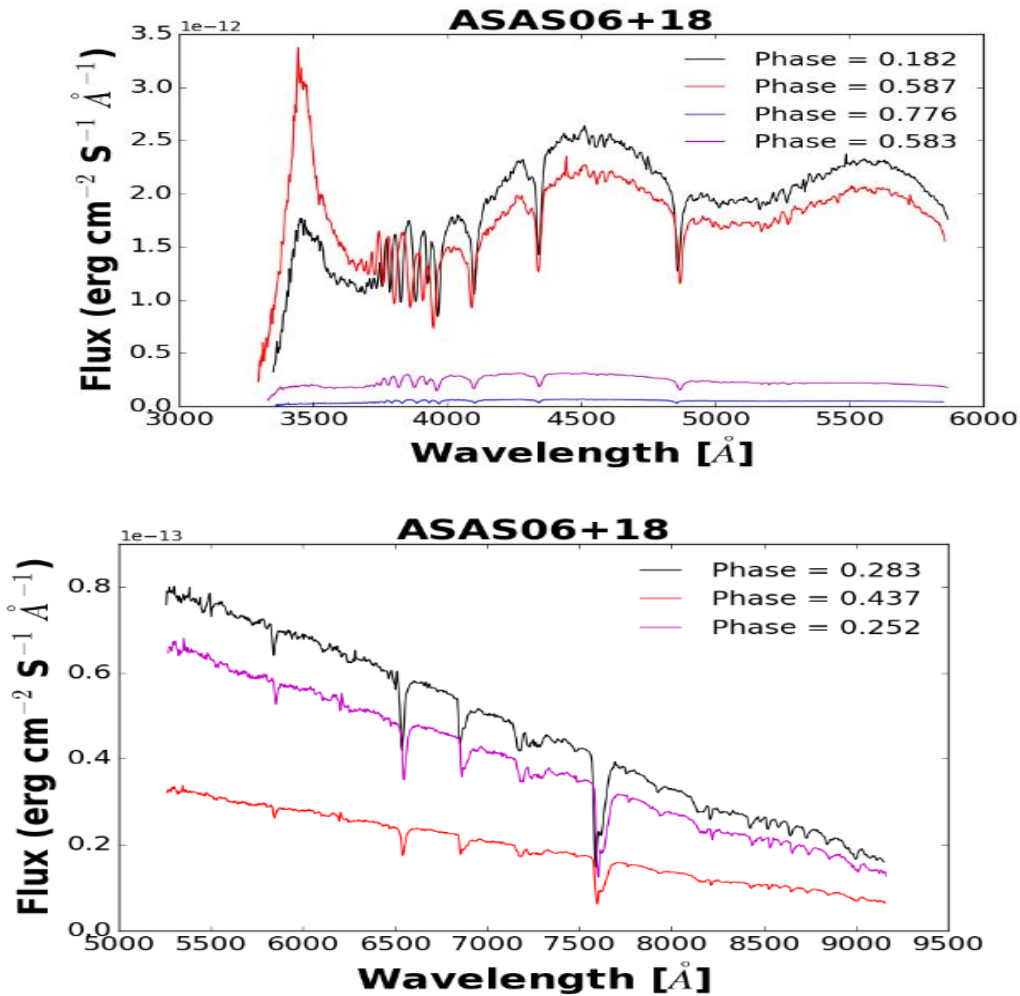

Figure 6: The observed spectra of ASAS06+18 in the blue (left panel) and red (right panel) bands at different phases.

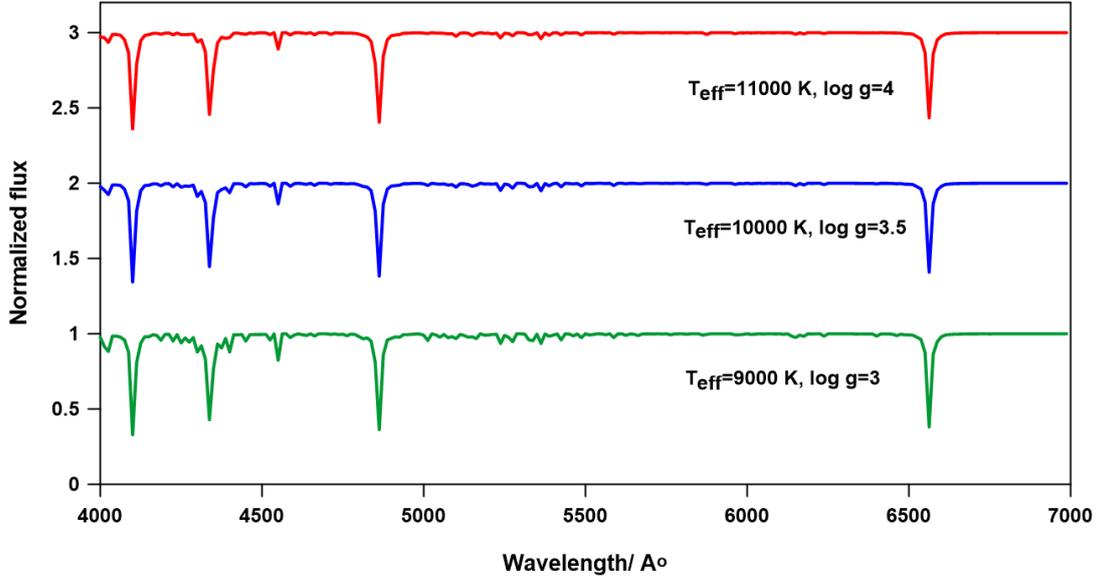

Figure 7: Normalized synthetic spectra of three models, labeled with effective temperature and surface gravities.

Fitting the observed $H_\alpha$, $H_\beta$, $H_\gamma$, $H_\zeta$ lines to the synthetic ones for the two phases 0.411 and 0.578 give $T_{eff} = 7000 \pm 250$ K, and $\log g = 4.0 \pm 0.2$ dex; $T_{eff} = 7250 \pm 250$ K, and $\log g = 4.0 \pm 0.2$ dex, respectively. We adopted the star's mean effective temperatures and surface gravities as $T_{eff} = 7125 \pm 250$ K and $\log g = 4.0 \pm 0.2$ dex. Figure 8 depicts the best fit of distinct phases of spectral lines in both the red and blue regions of the spectrum. In most cases, we could find a good match for the line centers and wings. This effective temperature is in good agreement with Qian et al. (2018), $T_{eff} = 7279 \pm 100$ K, while the surface gravity has a bit difference ($\log g = 3.89 \pm 0.13$ dex). We excluded the results from the fitting to the spectrum at other phases due to problems with the goodness of the spectrum, which led to high effective temperatures and surface gravities.

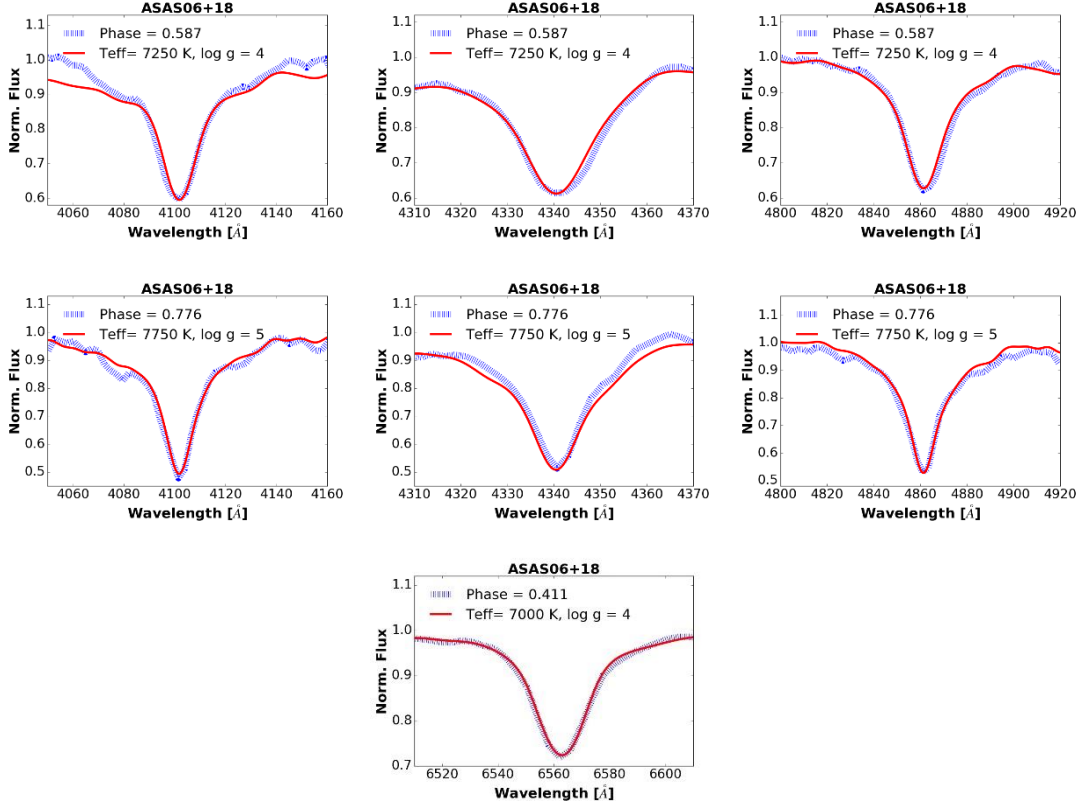

Figure 8: Comparison of observed Line profiles at different phases with that from synthetic spectra for the star ASAS06+18. Fitting line profiles at the two phases 0.411, 0.578, and 0.766 give $T_{eff} = 7000 \pm 250$ K, and $\log g = 4.0 \pm 0.2$ dex, $T_{eff} = 7250 \pm 250$ K, and $\log g = 4.0 \pm 0.2$ dex, and $T_{eff} = 7750 \pm 250$ K, and $\log g = 5.0 \pm 0.2$ dex, respectively.

## 4. Physical Parameters and Evolutionary State

The fitting of the spectra of ASAS06+18 with the synthetic ones gives the effective temperature and the surface gravity as $T_{eff} = 7125 \pm 250$ K, and $\log g = 4.0 \pm 0.2$ dex. To compute the remainder of physical parameters for the star, we adopted the parallax as 1.37±0.016 mas (Gaia DR1, 2016), corresponding to 729.9±9.0 pc if the reddening in this low galactic latitude area (b=4.293) is ignored. Mv=2.513±0.024 for the distance and Mv=2.456±0.024 for the mean V-brightness. The spectral type (or the mean *B-V*=0.33) implies a bolometric correction of BC=-0$^m$.0955±0.003 (Reed, 1998). The bolometric magnitude, $M_{bol}$, is further provided by $M_\lambda = M_{bol} - BC_\lambda$ and the star's absolute magnitude was determined in the visible filter as $M_{bol}$=2.36±0.046 using Reed (1998) and Zwintz et al. (2004). The (B-V)$_o$ is calculated using the equation

$$(B - V)_0 = -3.684 \, \log T_{eff} + 14.551, \quad \text{for } \log T_{eff} < 3.961.$$

The radius of the star is derived from a polynomial fit to the temperature/radius relation by Tsvetkov (1988) as $R/R_\odot$=1.626±0.077. The masses M could be computed using Cox (1989) relation as M =1.530±0.078 $M_\odot$, and the luminosity is calculated as L=5.714±1.066 $L_\odot$. The pulsation constant might be computed using the relationship

$$\log Q = 0.5 \log g + 0.1 M_{bol} + \log T_{eff} + \log P - 6.456,$$

which gives the Q value of f1 was 0.0338±0.0003 days, which was within the predicted range for the fundamental mode Fitch (1980). The adopted physical parameters of ASAS06+18 are listed in Table 5.

Table 5: Physical Parameters of ASAS06+18.

| $T_{eff}$ (K) | $M/M_\odot$ | $L/L_\odot$ | $R/R_\odot$ | $M_{bol}$ | BC | log g (dex) | Age x$10^9$(yr) | Q (days) |
|---|---|---|---|---|---|---|---|---|
| 7125± 250 | 1.530± 0.078 | 5.714± 1.066 | 1.626± 0.077 | 2.36± 0.016 | -0.0867 | 4.0±0.2 | 1.33 | 0.0338± 0.0003 |

In Figure 9, we displayed Girardi et al. (2000) mass-luminosity (M-L) and mass-radius (M-R) relations for both zero-age main-sequence stars (ZAMS) and terminal-age main-sequence stars (TAMS) with Z = 0.014. The ASAS06+18 locations on the two diagrams are exactly on the ZAMS tracks, indicating that it is an unevolved star. We depicted the position of ASAS06+18 on the log $T_{eff}$-log g graphs for the mass tracks suitable for the mass of ASAS06+18 (1.4 $M_\odot$ -2 $M_\odot$) in the lower-left panel, for the metallicity values Z=0.019 ([Fe/H] = -1.61) and Z=0.03 ([Fe/H] = -1.41). The computed mass (1.595 $M_\odot$) agrees with the evolutionary mass track with 1.6 $M_\odot$. We also plotted the isochrones suitable for the effective temperature and luminosity of ASAS06+18 in the lower right panel of Figure 9, which indicate that the star hasn't crossed the red border of the instability strip (RE) and is still in between both borders. The age of ASAS06+18 is calculated using fitting to isochrones is 1.33 x$10^9$ years.

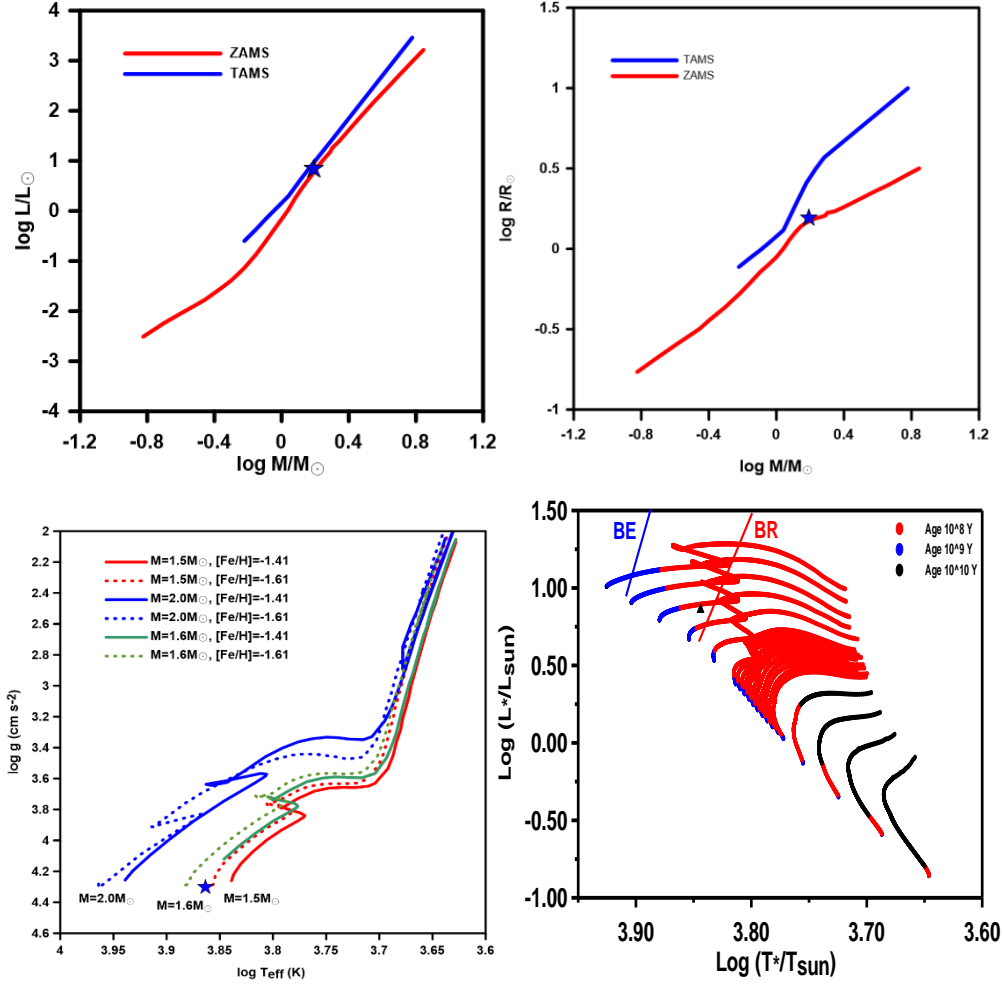

Figure 9: Positions of the star ASAS06+18 on the mass-luminosity (upper left panel) and mass-radius (upper right panel) diagrams of Girardi et al. (2000) evolution models. The temperature-gravity (lower left panel) and temperature-luminosity (lower right panel) of Girardi et al. (2000) evolution models for the metallicity Z=0.03 (solid lines) and Z=0.019 (dashed lines).

## 5. Discussion and Conclusion

We presented the first spectroscopic and photometric study of the new pulsating star ASAS06+18. The star was observed using the Kottamia Faint Imaging spectropolarimeter (KFISP) installed to the Cassegrain focus at Kottamia Astronomical Observatory (KAO) with two grisms encompassing the spectral regions $\lambda\lambda$ 3360-5870 Å and $\lambda\lambda$ 5300-9180 Å, with spectral resolutions of 1025 Å in blue and 1133 Å in red. Synthetic modeling of some line profiles low-resolution spectra at different phases of the star provided the effective temperature and surface gravity as $T_{eff} = 7125 \pm 250$ K, and $\log g = 4.0 \pm 0.2$ dex. Although our spectroscopic observations and that of Qian et al. (2018) are low-resolution (the resolving power of LAMOST spectra is ~1800), we think the present effective temperature and surface gravity are more precise because we calculated them as a mean of values computed at different phases.

The analysis of the BVR photometry revealed that the star is a δ Scuti star with low amplitude (less than 0.1 magnitudes in the *V*-band) and a short period (P=0.071253065 d). A newly updated frequency of 13.0035232 cd$^{-1}$ (0.076902235 d) with an amplitude of 49.93 (0.50) at phases 0.326 (0.002) and S/N 21.75 and Fourier analysis of the light curves reveals the basic model with two frequencies. Based on the available data, 37 new times of maximum light are obtained, as well as an updated ephemeris and O-C data for the star. Assuming its period decreases and changes gradually, we calculate the new value of $(1/P)\,dP/dt$ as $41.49579471\,yr^{-1}$. The amplitude of the B, V, and R filters are 0.075, 0.064, and 0.048 mags, respectively. These amplitude values indicate that ASAS06+18 is of the sub-type LADS.

The star's physical parameters are compared to the theoretical evolutionary model, where the star is located on the ZAMS and TAMS tracks of the M-L and M-R diagrams, indicating that the star appears to be unevolved. The present result agrees with the analysis of Bedding et al. (2020) applied to thousands of δ Scuti candidates; their results indicated that δ Scuti stars with consistent frequency spacings and masses ranging from 1.5 to 1.8 solar masses are found near the zero-age main sequence (ZAMS). The star's location on the $T_{eff}$-log g diagram shows good agreement between the adopted mass (M=1.595 $M_\odot$) and the evolutionary track with mass M=**1.6** $M_\odot$. We determined the star's age from the isochrones as 8.530x10$^8$yr.

We think that obtaining more photometric observations and high-resolution spectroscopic observations of the star may allow precise chemical analysis to search for the evidence of accretion from circumstellar material, calculate the effective temperature and surface gravity precisely, and perform an asteroseismology study. Besides the optical observations, we shall use TESS and Kepler observations in future work.

**Acknowledgments**

The research is supported by the Academy of Science Research and Technology, Egypt (ASRT), under the project titled "Investigating the Spectral and Photometric Behavior of the Cataclysmic Variables in Multi-Wavelength". This paper is based upon work supported by Science, Technology & Innovation Funding Authority (STDF) under grant number STDF 45779.